\documentclass[showpacs,prl,twocolumn,preprintnumbers,amsmath,amssymb,superscriptaddress]{revtex4}

\usepackage{amsmath,amssymb,graphicx,color,dcolumn,bm}
\usepackage[hypertex,bookmarks=true]{hyperref}
\usepackage{setspace,atbegshi,graphicx,color}
\def\bra#1{\langle #1|}
\def\ket#1{|#1\rangle}

\begin{document}

\title{Iyoda, Kaneko, and Sagawa Reply}

\author{Eiki Iyoda}
\affiliation{
Department of Applied Physics, The University of Tokyo,
7-3-1 Hongo, Bunkyo-ku, Tokyo 113-8656, Japan
}
\author{Kazuya Kaneko}
\affiliation{
Department of Applied Physics, The University of Tokyo,
7-3-1 Hongo, Bunkyo-ku, Tokyo 113-8656, Japan
}
\author{Takahiro Sagawa}
\affiliation{
Department of Applied Physics, The University of Tokyo,
7-3-1 Hongo, Bunkyo-ku, Tokyo 113-8656, Japan
}

\maketitle

Gemmer {\it et al.} in their Comment~\cite{Gemmer2017} make criticisms on our Letter~\cite{Iyoda2017}, while they agree that our result is mathematically sound. Their arguments are summarized as (i) the Lieb-Robinson (LR) time $\tau_{\mathrm{LR}}$ is too short and ``unphysical'' for some examples, (ii) the average entropy production $\langle\sigma\rangle$ becomes negative for some parameters, and (iii)  in our numerical simulation, the integral fluctuation theorem (IFT) $\langle e^{-\sigma}\rangle=1$ holds only while the system remains in its initial state. Here we discuss that (i) and (ii) are not justified, but that (iii) is indeed a subtle point because of the large finite-size effect.

\underline{Reply to (i).} 
We agree that the LR times  of  their examples (proteins in water and coffee in air) are very short compared to their time scales (Brownian motion and daily life).  However, these situations are clearly not relevant to our theory in \cite{Iyoda2017}, which is for isolated quantum systems where the bath is initially in a pure state (specifically in an energy eigenstate).
In fact, the setups discussed in \cite{Gemmer2017} are far from isolated and pure.
In other words, in their setups the fluctuation theorem is not emergent from quantum mechanics, but is simply a consequence of the conventional scenario based on classical stochastic dynamics.
Our result  in \cite{Iyoda2017}  implies that if air or water was initially in an energy eigenstate, then the IFT would hold only within such a very short time scale.

We  emphasize that our main focus is on ideally-isolated artificial quantum systems such as ultracold atoms, as explicitly stated in  \cite{Iyoda2017}.
In fact, the LR time is  reasonably long for ultracold atoms, compared to the time scale of real experiments.
For example, in a typical experiment \cite{Cheneau2012}, the experimental time scale is $\hbar/J$ and the LR time is  $\tau_\mathrm{LR}\sim l\hbar/J$, where $J$ is the tunneling amplitude and $l$ is the side length of ${\rm B}_1$ in~\cite{Iyoda2017}. 

\underline{Reply to (ii).} 
The reason why the authors of~\cite{Gemmer2017} observed the negative entropy production for some parameters is that their choice of the initial eigenstate is not thermal for such cases, as detailed below.
We note that our theory \cite{Iyoda2017}  states that the second law and the IFT hold if the initial eigenstate is thermal.

In the hard-core boson model in~\cite{Gemmer2017,Iyoda2017}, the total Hilbert space of bath B is divided into particle-number sectors, labeled by $N$.
An energy eigenstate is thermal only if  it is in a sector whose $N$ is close to the average particle number $N^\ast$ in the canonical ensemble, because the strong eigenstate thermalization hypothesis (ETH) is valid only within each particle number sector~\cite{DAlessio2016}.
We note that the weak ETH is true without dividing the energy shell to particle-number sectors.
Because $N^\ast \simeq 15.9$ with $\omega = -50$ and $\beta = 0.1$, the choice of $N=4$ in \cite{Gemmer2017}  is far from $N^\ast$, which makes the initial eigenstate athermal.

\begin{figure}[t]
\begin{center}
\includegraphics[width=\linewidth]{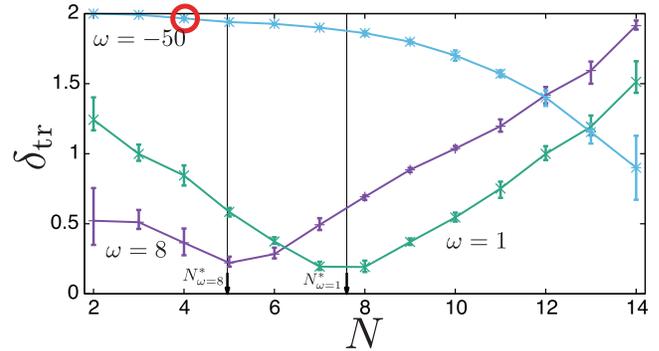}
\end{center}
\label{Main_fig1}
\caption{
The $N$-dependence of $\delta_\mathrm{tr}$. The Hamiltonian and the lattice are the same as in \cite{Gemmer2017,Iyoda2017}. 
The parameters are given by $\omega=1,8,-50$, $g=0.1$, and $\beta=0.1$.
$N^\ast_{\omega=1}$ and $N^\ast_{\omega=8}$ are the average particle numbers in the canonical ensemble for $\omega=1,8$, respectively.
For each data point, 10 energy eigenstates are sampled, and the error bar represents their standard deviation.
The red circle indicates the parameters used in \cite{Gemmer2017} ($\omega=-50$ and $N=4$). 
 }
\end{figure}

To directly show this, we calculated the trace norm between the reduced density operators of an energy eigenstate $\ket{E_i}$ and the corresponding canonical ensemble: $\delta_\mathrm{tr}:=\|\mathrm{tr}_{\mathrm{B}_2}[\ket{E_i}\bra{E_i}]-\mathrm{tr}_{\mathrm{B}_2}[\hat{\rho}_{\mathrm{B},\mathrm{can}}]\|_1$.
Figure 1 shows  the $N$-dependence of $\delta_\mathrm{tr}$, where we take $\mathrm{B}_1$ as the $2\times 2$ lower-left sites of bath B.
As shown in Fig.~1, $\delta_\mathrm{tr}$ takes a smaller value when $N$ is closer to $N^*$. For   $\omega=-50$ and $N=4$ (red circle), $\delta_\mathrm{tr}$ is large and $\ket{E_i}$ is not at all thermal. 

We have also confirmed that the entropy production is positive for a broad range of parameters, as long as the initial eigenstate is thermal.  We calculated $\langle \sigma \rangle$ at $t=\tau_{\rm LR}$ for $\omega = \pm 1, \pm 2, \pm 4, \pm 8, \pm 16, \pm 32, - 50$, $\beta =0.1,0.3$, $\gamma ' = 0.05,0.1,0.4,1.0,4.0$, $g=0.1,0.4$, and $N=4$. We found that $0.0076 \leq \langle \sigma \rangle \leq 1.84$ if $\delta_\mathrm{tr} < 0.3$ (thermal), while $-8.86 \leq \langle \sigma \rangle \leq 0.095$ if $\delta_\mathrm{tr} > 1.7$ (athermal).

\underline{Reply to (iii).}  
In~\cite{Iyoda2017}, we concluded that the IFT nontrivially holds in the short time regime based on the fact that the quantitative error evaluation, scaling of $\propto t^2$, is consistent with the prediction by the LR argument (the inset of Fig.~3 in \cite{Iyoda2017}).
On the other hand, the authors of \cite{Gemmer2017} argued that this scaling is just a general property of quantum systems.  
 
After careful consideration with some additional numerical simulation, we have to admit that it is hard to conclude whether the $t^2$-scaling comes from the LR argument or not based on our numerical data, because the finite-size effect is very large within the numerically-accessible system size.
We expect, however, that the IFT can be verified more clearly by using ultracold atoms, with which the system size can be much bigger than numerics.
We note that the time range where the IFT holds will linearly increase with $l$.
If one can take $l \sim 10^2$, the LR time becomes a hundred times of the experimental time scale, which can be realized with the current or the near-future technologies.

In addition, we remark that $|\gamma^\prime|$ should not be taken too large to verify the IFT, because the larger $|\gamma^\prime|$ is, the more the condition (8) in \cite{Iyoda2017} is violated.
In this respect, the emphasis in \cite{Gemmer2017} on the failure of the IFT for $\gamma^\prime = 4$ is misleading.

\begin{acknowledgments}
The authors thank Kyogo Kawaguchi and Takashi Mori for valuable discussions.
This work is supported by JSPS KAKENHI Grant Number JP16H02211, JP15K20944, and JP25103003.
\end{acknowledgments}


\end{document}